arXiv:1010.1112v1 [cs.DC] 6 Oct 2010

# Deterministic and Energy-Optimal Wireless Synchronization

Leonid Barenboim* Shlomi Dolev† Rafail Ostrovsky‡

November 8, 2018

**Abstract**

We consider the problem of clock synchronization in a wireless setting where processors must power-down their radios in order to save energy. In this setting, each processor has a radio device that is either on or off. When the radio device of a processor is on, it is able to communicate with other processors in its range. However, turning the radio on results in a significant waste of energy, even when listening. Energy efficiency is a central goal in wireless networks, especially if energy resources are severely limited. This is indeed the case in sensor networks, ad-hoc networks, and many other wireless network settings. Consequently, the main goal of multiple papers in wireless and sensor networks literature aims at achieving clock synchronization in an energy-efficient manner. In other words, the goal is to synchronize all clocks while minimizing the number of times a processor must switch its radio on.

The problem of clock synchronization is an important problem in the field of distributed algorithms. In the current setting, the problem is to synchronize clocks of $m$ processors that wake up in arbitrary time points, such that the maximum difference between wake up times is bounded by a positive integer $n$, where time intervals are appropriately discretized to allow communication of all processors that are awake in the same discrete time unit. (We remark that in this model we do not consider the issue of Broadcast Interference, which is a different problem known as radio broadcast problem.) The current model received a wide attention in sensor network literature. Currently, the best-known results for synchronization for single-hop networks of $m$ processors is a randomized algorithm due to Bradonjic, Kohler and Ostrovsky [2] of $O\left(\sqrt{n/m} \cdot \textit{poly-log}(n)\right)$ awake times per processor and a lower bound of $\Omega\left(\sqrt{n/m}\right)$ of the number of awake times needed per processor [2]. (Their algorithm synchronizes the $m$ processors with high probability, but may fail.) The main open question left in their work is to close the poly-log gap between the upper and the lower bound and to de-randomize their probabilistic construction and eliminate error probability. This is exactly what we do in this paper.

That is, we show a *deterministic* algorithm with radio use of $\Theta\left(\sqrt{n/m}\right)$ that never fails (and has a small hidden constant). We stress that our upper bound exactly matches the lower bound proven in [2], up to a small multiplicative constant. Therefore, our algorithm is *optimal* in terms of energy efficiency and completely resolves a long sequence of works in this area. In order to achieve these results we devise a novel *adaptive* technique that determines the times when devices power their radios on and off. Our novel technique may be of independent interest in other clock synchronization problems. While getting deterministic bound with extra logarithmic multiplicative term requires a new simple idea, getting rid of this additional term requires nontrivial combinatorial work. The contribution of this paper is both shwoing the simple new idea and then closing the gap to get exact upper and lower bounds.

In addition, we prove several lower bounds on the energy efficiency of algorithms for *multi-hop networks*. Specifically, we show that any algorithm for multi-hop networks must have radio use of $\Omega(\sqrt{n})$ per processor. Our lower bounds holds even for specific kinds of networks such as networks modeled by unit disk graphs and highly connected graphs. Our results imply that the simple deterministic algorithm devised for two-processor networks in Bradonjic et al. paper [2] with efficiency $O(\sqrt{n})$ can be used in multi-hop networks, and it is the most efficient solution in terms of energy use.

*Ben-Gurion University. Email: `leonidba@cs.bgu.ac.il`

†Ben-Gurion University, Email: `dolev@cs.bgu.ac.il`

‡UCLA. Email: `rafail@cs.ucla.edu`

**Topic classification:** distributed algorithms, network algorithms.

# 1 Introduction

***Problem description and motivation:*** In wireless networks in general, and in sensor and ad hoc networks in particular, minimizing energy consumption is a central goal. It is often the case that energy resources are very limited for such networks. Consider, for instance, a sensor network whose processors are fed by solar energy. In such cases devising energy efficient algorithms becomes crucial. A significant energy use of a processor takes place when its radio device is on. Then, it is able to communicate with other processors in its transmission range whose radio devices are also turned on. However, it wastes significantly more energy than it would waste if its radio device was turned off. For example, in typical sensor networks [26] listening to messages consumes roughly as much energy as fully utilizing the CPU, and transmitting consumes up to 2.5 times more energy. Moreover, if a processor runs in an idle mode, and its radio device is off, it consumes up to 100 times less energy than it would consume if its radio device was on. Therefore, the time that a processor can operate using an allocated energy resource largely depends on how often its radio is turned on.

Processors in a wireless network may wake up at somewhat different time points. For example, in the sensor network powered by solar energy, processors wake up in the morning when there is enough light projected on their solar cells. If the processors are spread over a broad area, then there is a difference in the wake up times. The processors' clocks start counting from zero upon wake up. Since there is a difference in wake up times, the clocks get out of synchronization. However, many network tasks require that all processors agree on a common time counting. In such tasks processors are required to communicate only in certain time points, and may be idle most of the time. If the clocks are not synchronized, a certain procedure has to be invoked by each processor in order to check the status of other processors. During this procedure processors may turn their radio on constantly, resulting in a major waste of energy. Therefore, clocks must be synchronized upon wakeup in order to save energy and to allow the execution of timely mannered tasks. The clock synchronization itself must be as efficient as possible in terms of energy use. It is desirable that among all possible strategies, each processor selects the strategy that minimizes its radio use. The *energy efficiency* of a processor is the number of time units in which its radio device is turned on.

In this paper we devise energy efficient clock synchronization algorithms. The goal of a clock synchronization algorithm is setting the logical clocks of all processors such that all processors show the same value at the same time. In order to achieve this goal, each processor executes an adaptive algorithm, which determines the time points (with respect to its local clock) in which the processor will turn its radio device on, for a fixed period of time. Once a processor's radio device is on, it is able to communicate with other processors in its range whose radio devices are also on at the same time interval. Based on the received information a processor can adjust its logical clock, and determine additional time intervals in which its radio device will be turned on. This process is repeated until all processors are synchronized.

***Our Results:*** We consider single-hop networks of $m$ processors, such that the maximum difference between processors wake up times is $n$. (Henceforth, *the uncertainty parameter*.) We devise several *deterministic* synchronization algorithms, the best of which has radio efficiency $O(\sqrt{n/m})$ per processor. Our results improve the previous state-of-the-art algorithms devised by Bradonjic et al. [2]. In [2] *randomized* algorithms for synchronization single-hop networks were devised, whose energy efficiency is $O(\sqrt{n/m} \cdot polylog(n))$ per processor. Therefore, our *deterministic* results improve the best previous *randomized* results by a polylogarithmic factor. Moreover, Bradonjic et al. proved lower bounds of $\Omega(\sqrt{n/m})$ per processor for the energy efficiency of any deterministic clock synchronization algorithm for single-hop networks. Hence, our algorithms are optimal in terms of radio use up to constant factors. In addition, our algorithms do not employ heavy machinery, as opposed to the algorithm of [2], that employs expanders and sophisticated probabilistic analysis. In contrast, we devise a combinatorial construction that quickly "spreads" processors' radio use approximately equally in time, which surprisingly allows them to synchronize more efficiently via chaining synchronization messages with each other.

We also prove lower bounds for *multi-hop* networks. We show that any deterministic synchronization algorithm for an $m$-processor multi-hop network must have total radio use $\Omega(m \cdot \sqrt{n})$. In [2] a simple deterministic algorithm for 2-processor network was devised with energy efficiency $O(\sqrt{n})$ per processor. This algorithms is extendable to $m$-vertex networks, in the sense that each processor learns the differences between its clock and the clocks of its neighbors. The total radio efficiency of the extended algorithm is $O(m \cdot \sqrt{n})$. As evident from our results it is far from optimal for single-hop networks. However, for multi-hop networks, its total radio efficiency $O(m \cdot \sqrt{n})$ is the best possible up to constant factors. Our lower bounds hold even for very specific network types such as unit disk graph and highly connected graphs.

***High-level ideas:*** In the synchronization algorithm for $m$ processors devised in [2] each processor determines by itself the time points in which it turns its radio on. The decision of a processor does not depend by any means on the decisions of other processors. Such a non-adaptive strategy makes the algorithm sub-optimal unless the number of processors is constant. Moreover, the decisions are made using randomization, and, consequently, the algorithm may fail. (However, the probability of failure is very low, since it is exponentially in $n$ close to zero.) In contrast, our algorithms are *deterministic* and *adaptive*. In our algorithms, periodically, each processor deterministically decides for the time points in the future in which it will turn its radio on. Each decision is made based on all the information the processor have learnt from communicating with other processors before the time of decision. Such a strategy decreases the number of redundant radio uses. In other words, the radio of a processor is used only if this processor is essential for synchronization, and no other processor can be used instead. Since all processors use this strategy, the radio use of each processor is as small as possible.

In the optimal case, a processor $i$, $i = 1, 2, .., m$, wakes up at global time $(i-1) \cdot \lfloor (n/m) \rfloor$. Each processor considers an (almost) exclusive time interval of length $O(n/m)$. In other words, it may turn its radio on only within the $O(n/m)$ first time units from wake up. The number of time units in which the radio is on is even smaller, specifically, $O(\sqrt{n/m})$. The sum of lengths of all considered intervals is therefore $O(n)$. All the considered intervals cover the entire time interval starting at the wake up of the earliest processor, and ending at the wake up of the latest one. Each processor has a time point in which it overlaps with the next processor, i.e., in which both processors turn the radio on. In the described case all processors are synchronized in a rely-race manner, where each processor is synchronized with the processor that wakes up immediately after it. However, in general, the processors wake up at arbitrarily global times in the range $[0, n]$. Therefore, there may be dense time intervals, in which many processors wake up, and sparse time intervals, in which few processors wake up, or even none at all. In this case a difficulty arises due to the need of synchronizing isolated intervals. We overcome this difficulty by devising a more sophisticated synchronization algorithm.

Let $V$ be an $m$-vertex set representing the processors of the network, and $E$ an initially empty edge set. Each time a pair of processors $u, v \in V$ communicate with each other, add the edge $(u, v)$ to $E$. Once the graph $\mathcal{G} = (V, E)$ becomes connected, all $m$ processors can be synchronized. Each time a processor turns its radio on, it communicates with other processors that also turn their radio on in the same time. Consequently, additional edges are added to $E$, and the graph $\mathcal{G}$ changes. In all time points the graph $\mathcal{G}$ consists of clusters. Initially, each vertex is a cluster, and clusters are merged as time passes. Each time a new cluster is formed, the clocks of the processors in the cluster are synchronized using our cluster-synchronization procedures. Next, each processor selects exclusive (with respect to other processors in the cluster) time points in the future in which its radio will be turned on. For a sufficient number of points, such a selection guarantees that one of the processors in the cluster will turn the radio on in the same time with another processor from another cluster. This results in merging of the clusters. Our algorithms cause all clusters to merge into a single unified cluster that contains all $m$ vertices very quickly.

***Related work:*** Clock synchronization is one of the most intensively studied and fundamentally important fields in distributed algorithms. [1, 3, 4, 5, 7, 8, 9, 10, 14, 17, 19, 20, 21, 22, 23, 24, 25, 27, 28, 29]. The aspect of energy efficiency of clock synchronization algorithms was concerned in most of these works.

In [24] Polastre et al. devised an algorithm with energy efficiency $O(n)$ per processors. Each processor simply turns its radio on for $n+1$ consecutive time units upon wake up. Since the maximum difference between wake up points is $n$, this guarantees that all processors are synchronized. More efficient solutions were devised by Palchaudhuri and Johnson [23], and by Moscibroda, Von Rickenbach, and Wattenhofer [20]. In these solutions, each processor turn its radio on for $O(\sqrt{n})$ time units that are randomly selected. Their correctness is based on the birthday paradox, according to which there exists a time point that is selected by two processor with high probability. In this time point both processors turn their radio on and are able to synchronize.

Recently Bradonjic et al. [2] devised deterministic algorithms for synchronizing two processors with efficiency $O(\sqrt{n})$. They also devised randomized algorithms for synchronizing $m$ processors with efficiency $O(\sqrt{n/m} \cdot polylog(n))$ per processor. The polylogarithmic factor in the latter efficiency bound depends on the probability of correctness, and grows as the probability grows. In addition, [2] also prove that any deterministic algorithm for synchronizing $m$ processors has energy efficiency $\Omega(\sqrt{n/m})$ per processor.

Additional synchronization problems that do not deal with energy consumption were studied in various threads of research. However, their description is beyond the scope of this paper. For more information see, for example, the surveys of Lenzen et al. [18], Sundararaman et al. [30], Sivrikaya and Yener [28].

***Structure of the paper:*** In Section 2 we describe the setting, building blocks and definitions used in our algorithms. Section 3 contains our synchronization algorithms. Section 4 contains the lower bounds.

# 2 Preliminaries

### 2.1 The Setting

We use the following abstract model of a wireless network. We remark that although this abstract model is quite strong, it is sufficiently expressive to capture a more general case as explained in Appendix B. Global time is expressed as a positive integer, and available for analysis purposes only. The network is modeled by an undirected $m$ vertex graph $G = (V, E)$. The processors of the network are represented by vertices in $V$, and enumerated by $1, 2, ..., m$. For each pair of processors $u$ and $v$ residing in the communication range of each other there is an edge $(u, v) \in E$. Communication is performed in discrete rounds. Specifically, time is partitioned into units of equal size, such that one time unit is sufficient for a transmitted message to arrive at its destination. (At the physical level this can be relaxed such that communication is possible if two processors turn their radio on during intervals that overlap for at least one time unit.) A processor wakes up in the beginning of a time unit, and its physical and logical clocks start counting from zero. The clocks of all processors tick with the same speed, and are incremented in the beginning of each new time unit. The wake up time of the processors, and, consequently, the clock values in a certain moment may differ. However, the maximum difference between the wake up times of any two processors is bounded by an integer $n$, which is known to all processors. (In other words, each processor wakes up with an integer shift in the range $\{0, 1, ...n\}$ from global time 0.) See Appendix B for a discussion on more general cases. Specifically, the wakeup shifts may be non-integers, and the clock speeds may somewhat differ, as long as the ratio of different speeds is bounded by a constant.

Each processor has a radio device that is either on or off during each time unit. If the radio device is off, its energy consumption is negligible. The energy efficiency of an algorithm is the number of time units during which the radio device of a processor is on. A pair of processors $(u, v) \in E$ are able to communicate in a certain time unit $t$ (with respect to global time) only if the radio devices of both $u$ and $v$ are turned on during this time unit.

### 2.2 Algorithm Representation

The running time $f(n, m)$ of an algorithm is the worst case number of time units that pass from wake up until the algorithm terminates. The algorithm specifies initial fixed time points for a processor to turn its radio on. In addition, it adaptively determines new time points each time a processor turns its radio on. The time points are determined by assigning strings to processors as follows. The strings of the $m$ processors are represented using a two dimensional array $A$. The array $A$ contains $m$ rows. For $i = 1, 2, ..., m$, The $i$th row belongs to the $i$th processor. The number of columns of $A$ is $n + f(n, m)$. All cells of $A$ are set to 0, except the cells that are explicitly set to 1. (Initially, all cells are set to 0.) The algorithm specifies an initial fixed string $S_i$ for each processor $i$. For $i = 1, 2, ..., m$, suppose that processor $i$ wakes up at time $t_i$, $0 \leq t_i \leq n$, with respect to global time. Then the $i$th row of $A$ is initialized as follows. For $j = 0, 1, ..., |S_i| - 1$, set $A[i][t_i + j] = S_i[j]$. See Figure 1 (a) below.

For $j = 0, 1, 2...$, at local time $j$, a processor $i$ accesses the cell $A[i][t_i + j]$. A processor $i$ turns its radio device on at local time $k \geq 0$ if and only if $A[i][t_i + k] = 1$. If at global time $t$ the radio device of a processor $i$ is on, then it can communicate with all processors $j$ in its communication range for which $A[j][t] = 1$. Based on the received information, processor $i$ deterministically decides whether to update cells in the row $A[i]$. It can update, however, only cells that represent time points in the future, i.e., cells $A[i][t']$, for $t' > t$. Observe, however, that processor $i$ is unaware both of global time and the shift $t_i$. (In particular, it is unaware of the index of the cell it is accessing in the row $A[i]$.) The algorithm terminates once all processors detect a column of ones, i.e., a column $\ell$ such that for all $1 \leq j \leq m$, it holds that $A[j][\ell] = 1$. (Once all processors detect a column of ones, they all turn their radio on in the same time, and synchronize their clocks.) A clock synchronization algorithm $\mathcal{A}$ is correct if for all $i = 1, 2, ..., m$, for all shifts $t_i$, $t_i \in \{0, 1, 2, ..., n\}$, once $\mathcal{A}$ is executed by all processors there exists a column $\ell$ such that for all $j = 1, 2, .., m$, $A[j][\ell] = 1$. See Figure 1 (b).

**a)** A

| | | | | | | | | | | | | | | | | | | | | | | | | | | | | | | | | | | | | |
|---|---|---|---|---|---|---|---|---|---|---|---|---|---|---|---|---|---|---|---|---|---|---|---|---|---|---|---|---|---|---|---|---|---|---|---|---|
| 0 | 0 | 1 | 1 | 0 | 1 | 0 | 1 | 0 | 0 | 0 | 0 | 0 | 0 | 0 | 0 | 0 | 0 | 0 | 0 | 0 | 0 | 0 | 0 | 0 | 0 | 0 | 0 | 0 | 0 | 0 | 0 | 0 | 0 | 0 | 0 | 0 |
| 0 | 1 | 1 | 0 | 1 | 0 | 1 | 0 | 0 | 0 | 0 | 0 | 0 | 0 | 0 | 0 | 0 | 0 | 0 | 0 | 0 | 0 | 0 | 0 | 0 | 0 | 0 | 0 | 0 | 0 | 0 | 0 | 0 | 0 | 0 | 0 | 0 |
| 0 | 0 | 0 | 0 | 0 | 0 | 0 | 0 | 1 | 1 | 0 | 1 | 0 | 1 | 0 | 0 | 0 | 0 | 0 | 0 | 0 | 0 | 0 | 0 | 0 | 0 | 0 | 0 | 0 | 0 | 0 | 0 | 0 | 0 | 0 | 0 | 0 |

**b)** A

| | | | | | | | | | | | | | | | | | | | | | | | | | | | | | | | | | | | | |
|---|---|---|---|---|---|---|---|---|---|---|---|---|---|---|---|---|---|---|---|---|---|---|---|---|---|---|---|---|---|---|---|---|---|---|---|---|
| 0 | 0 | 1 | 1 | 0 | 1 | 0 | 1 | 0 | 0 | 0 | 0 | 0 | 0 | 0 | 0 | 0 | 0 | 0 | 0 | 0 | 0 | 1 | 1 | 0 | 1 | 0 | 1 | 0 | 0 | 0 | 0 | 0 | 0 | 0 | 0 | 1 |
| 0 | 1 | 1 | 0 | 1 | 0 | 1 | 0 | 0 | 0 | 0 | 0 | 0 | 0 | 0 | 0 | 0 | 0 | 1 | 1 | 0 | 1 | 0 | 1 | 0 | 0 | 0 | 0 | 0 | 0 | 0 | 0 | 0 | 0 | 0 | 0 | 1 |
| 0 | 0 | 0 | 0 | 0 | 0 | 0 | 0 | 1 | 1 | 0 | 1 | 0 | 1 | 0 | 0 | 0 | 0 | 0 | 0 | 0 | 0 | 0 | 0 | 0 | 0 | 1 | 1 | 0 | 1 | 0 | 1 | 0 | 0 | 0 | 0 | 1 |
| 0 | 1 | 2 | 3 | 4 | 5 | 6 | 7 | 8 | 9 | 10 | 11 | 12 | 13 | 14 | 15 | 16 | 17 | 18 | 19 | 20 | 21 | 22 | 23 | 24 | 25 | 26 | 27 | 28 | 29 | 30 | 31 | 32 | 33 | 34 | 35 | 36 |

**Fig. 1.** *Example of the array $A$ of three processors executing an algorithm with shifts $t_1 = 2, t_2 = 1, t_3 = 8$. (a) The array $A$ is initialized with the strings $S_1 = S_2 = S_3 = '110101'$. (b) The array $A$ after the termination of the execution.*

### 2.3 Building Blocks and Definitions

A *radio use policy* is a protocol for a processor $i \in \{1, 2, ..., m\}$ that determines the local time points in which the processor $i$ turns its radio on. For $r = 0, 1, 2, ...$, in the beginning of time unit $r$ from wakeup, the processor $i$ decides whether to turn its radio device on as explained above [1].

For a fixed string $s$ over the alphabet $\{0, 1\}$ and a positive integer $t$, an *$(s, t)$-radio-use policy* of a processor $i$ determines the local time units in which $i$ turns its radio on. For a processor $i$ that wakes up at global time $t_i$, we say that processor $i$ *performs an (s,t)-radio use policy* if it sets $A[i][t_i + t + j] = s[j]$, for $j = 0, 1, ..., |s| - 1$, and turns its radio device on accordingly. (Recall that processor $i$ turns its radio device on at local tick $k$ if and only if $A[i][t_i + k] = 1$.) The processor starts performing the policy at global time $t_i + t$. It completes the policy at global time $t_i + t + |s| - 1$. During this period a processor may select new time points in the future in which additional policies will be performed.

[1] The decision process can also be performed using a decision tree.

Next, we define the notion of *length*, *covering-weight*, and *covering-density* of a policy. These definitions are used in the correctness analysis of the algorithms.
The *length* of an $(s,t)$-radio-use policy $p$, denoted $len(p)$, is the difference between the positions of the first and last '1' in $s$ plus one. (In other words, if $j$ is the smallest index such that $s[j] = 1$, and $k$ is the largest index such that $s[k] = 1$, then $len(p) = k - j + 1$.) Intuitively, the length of a policy is the time duration required for performing the policy. For the $(s,t)$-radio-use policy $p$, the string $s$ is a concatenation of two substring $s' \circ s''$, defined by $p$. The substring $s'$ is called the *initial part* of $s$, and the substring $s''$ is called the *main part* of $s$. We say that *$i$ performs the initial part of $p$* in global time $t'$ if it performs the policy $p$, and the global time $t'$ satisfies $t + t_i \leq t' \leq t + t_i + |s'| - 1$. If $i$ performs the policy $p$, and the global time $t'$ satisfies $t + t_i + |s'| - 1 < t'$, we say that *$i$ performs the main part of $p$*.

We say that two processors $i$ and $j$ *overlap* if there is a global time point $t'$ in which both processors turn their radio on. Two processors $u$ and $v$ can communicate (either directly or indirectly) if they overlap, or if there exist a series of processors $w_1, w_2, ..., w_k$, $w_1 = u, w_k = v$ such that $w_i$ overlaps with $w_{i+1}$, for $i = 1, 2, ..., k - 1$. If such a series does not exist, we say that there is a point of discontinuity between $u$ and $w$. A *point of discontinuity* is a global time point $t'$ in which either (1) there is no processor that performs a radio use policy, or (2) each processor that do perform a radio use policy, completes it in time $t'$. A global time interval $(s', t')$ is *continuous* if there are no points of discontinuity in it. For a continuous interval $(s', t')$ such that $s'$ and $t'$ are discontinuity points, all processors performing a radio use policy during the interval $(s', t')$ form a *cluster c*. In this case we say that $c$ *covers* the interval $(s', t')$. The *length* of a cluster $c$ that covers an interval $(s', t')$, denoted $len(c)$, is $t' - s' + 1'$.

Each processor in a cluster adds weight to the cluster. Consequently, clusters containing many processors are heavier than clusters containing few processors. The *covering-weight* of a cluster $c$, denoted $cwet(c)$, is the sum of lengths of policies of processors contained in $c$. Consider two clusters $c$ and $c$ with the same covering-weight, but such that the length of $c$ is much shorter then the length of $c'$. Therefore, $c'$ covers a much longer time interval. We show later in this paper that clusters that cover longer intervals are 'better' in a certain way. Consequently, $c'$ is better than $c$, although they have the same covering weight. On the other hand, a short and light cluster may be better than a long and heavy one. Therefore, neither the length nor the covering-weight of a cluster are expressive enough to determine how 'good' a cluster is. Hence, we add the notion of covering-density, which is the ratio between covering-weight and length of a cluster. The *covering-density* of a cluster $c$, denoted $cden(c)$, is $\frac{cwet(c)}{len(c)}$. Clusters of lower covering-density are considered as better clusters. (Observe that these definitions are different from the usual definitions of string weight and density in which only the number of ones in the string are counted.)

Next, we give similar definitions for intervals. The *length* of an interval $q = (s', t')$, denoted $len(q)$, is $t' - s' + 1'$. Suppose that during interval $q$ there are $\ell$ policies that are performed. (Possibly, some have started before time $s'$, and some have ended after time $t'$.) Let $q_1, q_2, ..., q_\ell$ be the intervals contained in $q$ in which the main parts of the policies are performed. The *covering-weight* of an interval $q$, denoted $cwet(q)$, is $\Sigma_{i=1}^{\ell} len(q_i)$. The *covering-density* of the interval, denoted $cden(q)$, is $\frac{cwet(q)}{len(q)}$.

# 3 Synchronization Algorithms for Single-Hop Networks

## 3.1 Procedure Synchronize

In this section we present a *deterministic* synchronization algorithm for complete graphs on $m$ vertices with energy efficiency $O((\sqrt{n/m}) \log n)$ per processor. In the next section we devise an algorithm with energy efficiency $O(\sqrt{n/m})$ per processor. This result is optimal up to constant factors, as evident from the matching lower bound $\Omega(\sqrt{n/m})$ [2]. As a first step, we define the following basic radio use policy for a processor, consisting of two parts. Starting from local time $t$, For a given integer $k > 0$, turn the radio devise on for $k$ consecutive time units. (Henceforth, *initial part.*) Then, for the following $k^2$ time units, turn the radio on only once in each $k$ consecutive time units. (Henceforth, *main part.*) In other words, starting from the beginning of the main part, the radio is turned on during time units $k, 2k, 3k, ..., k^2$. This completes the description of the policy. It henceforth will be referred as *$k$-basic policy.* Its pseudocode is given below. The string $s$ of the policy is defined by $s[i] = 1, s[(i+2) \cdot k - 1] = 1$ for $0 \le i < k$. The length of a $k$-basic policy is $k + k^2$, but the number of time units in which the radio is used is only $2k$. We remark that the $k$-basic policy is defined for any positive integer $k$, but our algorithms employ policies in which $k = \Theta(\sqrt{n/m})$.

**Algorithm 1** Procedure Basic-Policy($k$,$T_v$)

A $k$-basic policy for a processor $v$ starting from local time point $T_v$

**for** $i := 0, 1, 2, ..., k^2 + k - 1$ **do**
    $s[i] := 0$
**end for**
**for** $i := 0, 1, 2, ..., k - 1$ **do**
    $s[i] := 1$
    $s[(i+2) \cdot k - 1] := 1$
**end for**
**for** each local time unit $T := T_v, T_v + 1, ..., T_v + k^2 + k - 1$ **do**
    turn radio in time unit $T$ if and only if $s[T - T_v] = 1$
**end for**

Consider a pair of processors $u$ and $v$ that wake up at the beginning of global time units $t_u$ and $t_v$, respectively, such that $t_u < t_v$. Suppose that both processors use the $k$-basic policy $p$ upon wakeup, and that $t_v - t_u < len(p)$. Then, there is a global time unit $t$ in which both processors turn their radio devices on. In this case we say that the processors *overlap.* (See Figure 2 below.) We summarize this fact in the next lemma.

**Lemma 3.1.** *Suppose that processors $u$ and $v$ wake up at global time points $t_u < t_v$, such that $t_v - t_u < len(p)$, and execute the $k$-basic policy $p$ upon wake up, for an integer $k > 0$. Then $u$ and $v$ overlap.*

*Proof.* We prove that the overlap occurs during the initial part of the policy performed by $v$. If $t_v - t_u < k$, then at global time $t_v$ less than $k$ time units have passed from the wake up times of both processors. Hence the overlap occurs at time $t_v$, since both processors turn their radio on for $k$ consecutive time units upon wake up. Otherwise, $k \le t_v - t_u < len(p)$. Since $u$ turns its radio on in global time $t_u + len(p) - 1$, there exist a global time point $t \ge t_v$ in which $u$ turns its radio on. Let $t'$ be the smallest integer such that $t' \ge t_v$, and $u$ turns its radio on in global time $t'$. Observe that according to the $k$-basic policy it holds that $t_v \le t' < t_v + k$, since during the policy execution there are no $k$ consecutive time points in which $u$ does not turn the radio on. Since the processor $v$ turns its radio on at global times $t_v, t_v + 1, ..., t_v + k - 1$, the processors $u$ and $v$ overlap at time $t'$. □

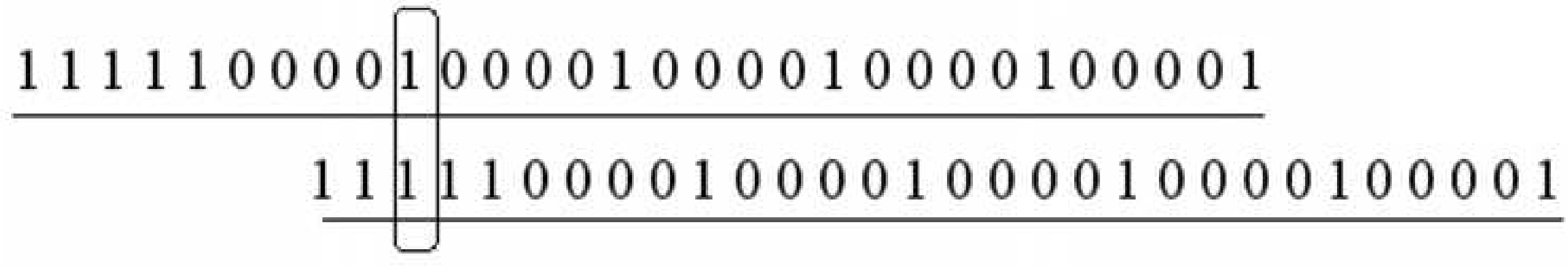

**Fig. 2.** *Two processors perform the 5-basic policy, and overlap.*

We say that processors *synchronize* their clocks if after the time point of synchronization the logical clocks of the processors show the same value at the same time. Any two overlapping processors synchronize their clocks as follows. Each processor executes the following procedure called *Procedure Early-Synch.* During its execution the processor that began performing its radio policy later among the two is synchronized with the other processor. In other words, the later processor updates its logical clock to be equal to the logical clock of the earlier processor. (Observe that the clock value of the later processor is not greater than that of the earlier processor, therefore, clocks do not go backwards.) To this end, each processor maintains the local variables $id, \tau, J$, where $id$ is the unique identity number of the processor, $\tau$ is the local clock value, and $J$ is the number of time units passed since the processor began performing the current radio policy. The variable $\tau$ is updated in each time unit by reading the logical clock value and assigning it to $\tau$. The variable $J$ is set to 0 each time the processor starts a radio use policy, and is incremented in each time unit. Each time a processor turns its radio on, it transmits the message $(id, \tau, J)$. Once a processor $u$ receives a message $(id_v, \tau_v, J_v)$ form a processor $v$, processor $u$ determines whether it began its radio policy after $v$ did. If so, $u$ updates its local clock to $\tau_v$. If both processors began their policy at the same time, then the clocks are synchronized to the clock of the processor with the greater Id. This completes the description of the Procedure. The pseudocode is given below. The next lemma states its correctness.

**Algorithm 2** Procedure Early-Sync()

A protocol for vertex $u$ that performs a policy, for any local round $r$.
Initially, $J_u = 0$.

$\tau_u :=$ local clock value of $u$
**if** radio device is on **then**
    send the message $(id_u, \tau_u, J_u)$
**end if**
**if** received a message $(id_v, \tau_v, J_v)$ **then**
    **if** $(J_u < J_v)$ or $(J_u = J_v$ and $id_u < id_v)$ **then**
        set local clock to $\tau_v$
        $J_u := J_v$
    **end if**
**end if**
$J_u := J_u + 1$

Return $J_u$ once the policy is completed.

**Lemma 3.2.** *Procedure Early-Synch executed by two overlapping processors synchronize their clocks.*

*Proof.* Recall that $t_v$ and $t_u$ are the global time points in which $u$ and $v$ begin performing their radio use policies, respectively, and that $0 < t_v - t_u < len(p)$. The correctness of the procedure follows from the fact that both processors $u$ and $v$ overlap. If $t_v - t_u < k$, then the overlap occurs at global time $t_v$. Otherwise, observe that any time interval containing the initial part of the policy of $v$ is of length $k$. It overlaps with $u$, since in the radio policy of $u$ each sequence of zeroes is of length $k-1$ followed by an occurrence of 1. Hence, there is a global time unit $t$ in which the radio devices of both processors are turned on. In this time unit each processor receives a message from the other one. Suppose that in time $t$ the processor $v$ receives the message $(Id_u, \tau_u, J_u)$ from $u$, and the processor $u$ receives the message $(Id_v, \tau_v, J_v)$ from $v$. Then exactly one processor updates its clock to the clock value of the other processor. Specifically, if $(J_u < J_v)$ or ($J_u = J_v$ and $id_u < id_v$), then $u$ updates its clock to $\tau_v$, and the clock value of $v$ remains $\tau_v$. Otherwise, $v$ updates its clock value to $\tau_u$, and the clock value of $u$ remains $\tau_u$. □

Procedure Early-Synch can be generalized for synchronizing a cluster containing an arbitrary number of processors. Recall that all processors in the cluster perform their policies in a time interval $(s', t')$ containing no discontinuity points. Hence, a message from a processor $u$ can be delivered to all processors that begin performing their policy after $u$ does so. The message is received directly by all processors that overlap with $u$, and is propagated in a rely-race manner to other processors. In this way all the processors in the cluster can be synchronized with the processor that was the first to start performing its policy.

The generalized procedure is called *Procedure Cluster-Synch*. During its execution all processors $u \in V$ perform the $k$-basic policy. A vertex $u$ starts performing its policy at local time point $T_u$ that is passed to the procedure as an argument. (The argument is passed by another procedure that invokes Procedure Cluster-Synch, which is described later in this section.) Each processor $u$ initializes a counter $J_u$ that is set to 0 once the policy starts, and is incremented by 1 in each time unit. Recall that the local clock of $u$ is represented by the variable $\tau_u$. Each time a radio device of a processor $u$ is on it transmits the message $(id_u, \tau_u, J_u)$. For each received message $(id_v, \tau_v, J_v)$ from a vertex $v$, if $(J_u < J_v)$ or ($J_u = J_v$ and $id_u < id_v$), then $u$ updates its clock to $\tau_v$ and its counter $J_u$ to $J_v$. This completes the description of Procedure Cluster-Synch. It pseudocode is provided below. Its correctness is proven in Lemma 3.3. It follows from the observation that all processors eventually synchronize their counters $J$ with the counter of the earliest processor in the cluster.

**Algorithm 3** Procedure Cluster-Synch($T_u$,$k$)

An algorithm for processor $u$.

```
Perform the k-basic policy starting from local time T_u
J := Early-Synch()
return J
```

**Lemma 3.3.** *For a fixed $k > 0$, suppose that processors $v_1, v_2, ..., v_\ell$ perform Procedure Cluster-Synch($T_{v_i}, k$), with the parameters $T_{v_1}, T_{v_2}, ..., T_{v_\ell}$, respectively. If in the resulting execution the processors $v_1, v_2, ..., v_\ell$ form a cluster, then $v_1, v_2, ..., v_\ell$ synchronize their clocks to the clock of the earliest processor $v_1$.*

*Proof.* Let $t_1, t_2, ..., t_\ell$ be the global times in which the processors $v_1, v_2, ..., v_\ell$ begin performing their policy, respectively. Assume without loss of generality that $t_1 \le t_2 \le ... \le t_\ell$. Assume also that $v_1$ is the processor with the greatest Id among the processors $v_j$ with $t_j = t_1$. We prove by induction on $i = 1, 2, ..., \ell$ that a processor $v_i$ is synchronized with the earliest processor $v_1$ once $v_i$ completes the initial part of its policy.

**Base case ($i = 2$):** Observe that $v_1$ and $v_2$ overlap, since there are no points of discontinuity in the cluster. The overlap occurs during the execution of the initial part of the policy of $v_2$. Therefore, once $v_2$ completes the initial part of its policy, it is synchronized with $v_1$.

**Induction step:** Suppose that once $v_{i-1}$ completes the initial part of its policy it is synchronized with $v_1$. Since there are no points of discontinuity, the processors $v_{i-1}$ and $v_i$ overlap. Let $t$ be the last global time point in which $v_{i-1}$ performs the initial part of its policy. (In other words, $v_{i-1}$ completes the initial part of its policy at time $t$.) The last time point $t'$ in which $v_{i-1}$ and $v_i$ overlap occurs once $v_{i-1}$ completes the initial part of its policy, or later. Hence, $t \leq t'$. By the induction hypothesis, at time $t$ the processor $v_{i-1}$ is synchronized with $v_1$. From this point and on it remains synchronized with $v_1$. Hence, at time $t' \geq t$, the processor $v_i$ receives a message with the clock value of $v_1$ and updates its clock accordingly. □

Next, we consider the most general problem in which $m$ processors wakeup at arbitrary global time points in the time interval $[0, n]$. If each processor performs the $k$-basic policy upon wakeup, then several clusters may be produced. The processors in each cluster can be synchronized using procedure Cluster-Synch. However, the execution of procedure Cluster-Synch will not synchronize processors from distinct clusters since any two distinct clusters are separated by a discontinuity point. We devise a procedure, called *Procedure Synchronize* that merges these clusters gradually, until only a single cluster remains. To this end, the parameter $k$ is selected to be large enough to guarantee that certain clusters have large covering-density. The processors in a cluster with large covering-density schedule the next policy execution times in a specific way that enlarges the length of the cluster to the maximal extent. Somewhat informally, the cluster is extended roughly equally to both of its sides. In other words, there is an integer $L > 0$ such that in the next phase the cluster begins $L$ time units earlier than in the previous phase, and terminates $L$ units later than in the previous phase. For a precise definition see Algorithm 4, line 14. The extension of the cluster to both of its sides prevents time drifts, and, consequently, in each phase some clusters overlap. Overlapping clusters are merged into fewer clusters of greater covering-weight.

**Algorithm 4** Procedure Flatten($J_u$, $k$)

A protocol for a vertex $u$, executed once $u$ completes the initial part of its policy

1: /*** First stage ***/
2: $J := J_u$
3: wait for $2n - J$ time units
4: /*** Second stage ***/
5: $B := \{(Id_u, J)\}$
6: transmit ($Id_u$,$J$)
7: **for** each received message $m = (Id_v\ ,J')$ **do**
8: $\quad B := B \cup \{m\}$
9: **end for**
10: $B'$ := sort $B$ by Ids in ascending order
11: $len(c) := (\max\{J' | (Id, J') \in B'\}\ )$
12: $\mu$ := the position of $(Id_u, J)$ in $B'$
13: $\ell := |B|$
14: $next := \left\lfloor 2n + \tau_u + \frac{len(c) - \ell \cdot k^2}{2} + \mu \cdot k^2 \right\rfloor$
15: return $next$ /* returned locally to the caller of this procedure */

The procedure for extending the length of a cluster is called *Procedure Flatten*. It is executed by processors in a synchronized cluster $c$, and proceeds in two stages. The first stage (See Algorithm 4, lines 1-3) is executed by each processor $u$ in the cluster once the processor $u$ is synchronized with the first processor of the cluster $v_1$. (In other words, once $u$ completes the initial part of its $k$-basic policy.) Then the counters $J$ of $u$ and $v_1$ are also synchronized. A processor $u$ schedules the next execution of its policy to be executed in $2n - J$ time units. The second stage (Algorithm 4, lines 4-15) is executed once $u$ performs the policy the next time. Observe that it is executed in the same time by all processors in the cluster. All processors of the cluster turn their radio on and learn the number of processors in the cluster, their ids, and the length of the cluster $len(c)$ with respect to the first stage. (We describe how to determine $len(c)$ shortly.) Each processor sorts the ids and finds its position $\mu$ in the sorting. If the current local time is $\tau$, and the number of processors in the cluster is $\ell$, it schedules the next policy execution to local time $\left\lfloor 2n + \tau + \frac{len(c) - \ell \cdot k^2}{2} + \mu \cdot k^2 \right\rfloor$, and returns this value.

The length of the cluster $len(c)$ is equal to the difference between the global time points of the beginning and the end of the cluster. Therefore, the length $len(c)$ is determined by the latest processor in $c$. Once the latest processor $v_\ell$ completes its policy in the first stage, its counter $J_\ell$ (which is synchronized with the counter of the earliest processor) is equal to the number of time units passed since the cluster has started. Once $v_\ell$ completes its policy, the entire cluster $c$ is completed. Hence, at that moment, it holds that $len(c) = J_\ell$. All processors learn this value in the second stage. (See step 11 in the pseudocode of Algorithm 4.) This completes the description of the procedure. Its properties are summarized below. See Figure 3 for an illustration.

**Lemma 3.4.** *Suppose that Procedure Flatten is executed by a cluster $c$ of $\ell$ processors that is formed in a global time interval $[p, q]$. Then*
*(1) The second stage of Procedure Flatten is performed at global time $p + 2n$ by all processors of $c$.*
*(2) Performing the policies by the scheduling of the second stage forms a cluster $c'$ of length $\ell \cdot k^2$.*
*(3) The cluster $c'$ covers an interval that contains the interval $[4n + \frac{p+q}{2} - \frac{\ell \cdot k^2}{2}, 4n + \frac{p+q}{2} + \frac{\ell \cdot k^2}{2}]$.*

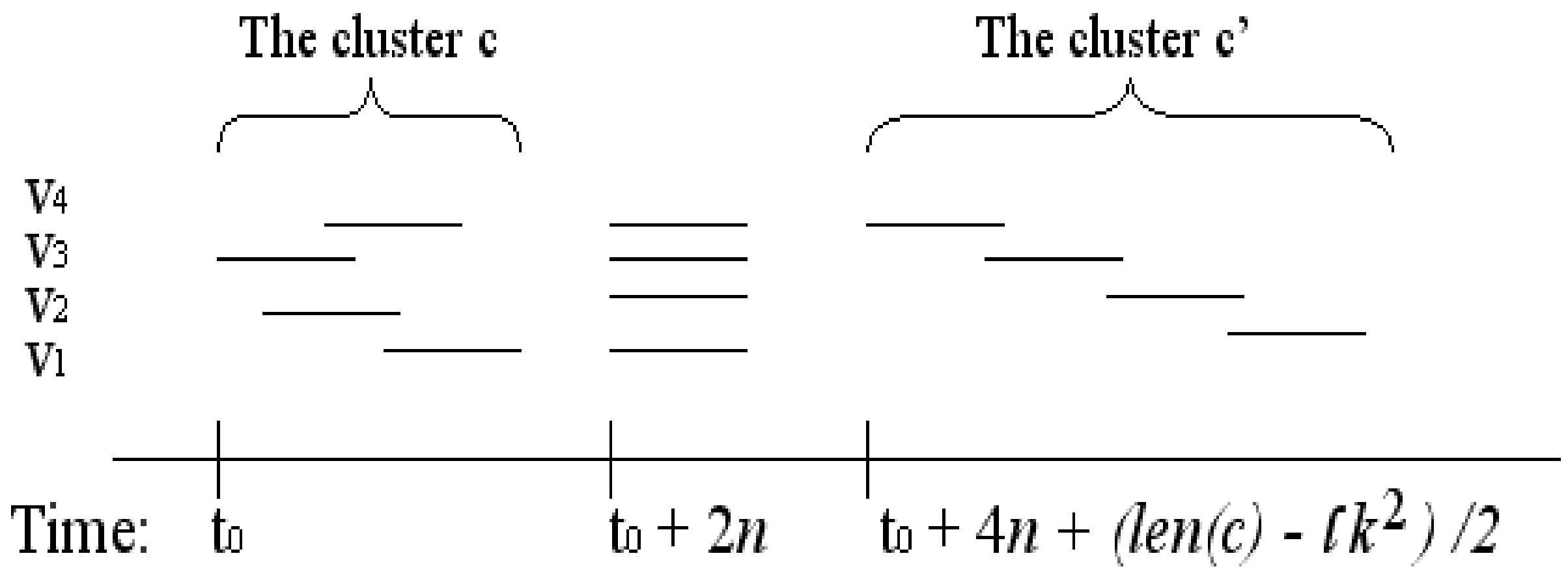

**Fig. 3.** *Illustration of Procedure Flatten with four processors. ($\ell = 4$.)* [1]

In order to synchronize $m$ processors that wake up at arbitrary times from the interval $[0, n]$, set $k = \left\lceil \sqrt{8 \cdot n/m} \right\rceil$. Procedure Synchronize is performed in phases as follows. For $i = 1, 2...$, the $i$th phase starts in global time $(i-1) \cdot 4n$. In each phase, two iterations are performed. Initially, in the first iteration of the first phase, each processor performs the $k$-basic radio policy upon wake up. Consequently, clusters are formed in the interval $[0, 2n]$. Each cluster is synchronized using Procedure Cluster-Synch. In the second iteration of the first phase, Procedure Flatten is performed. Then the next phase starts. In the first iteration of each phase, the $k$-basic policy is performed by each processor starting from a time point that was scheduled for it in the previous phase by Procedure Flatten. Consequently, new clusters are formed and synchronized. In the second iteration, Procedure Flatten is performed, and schedulings for the next phase are determined. Procedure Synchronize terminates once the interval $[i \cdot 4n, i \cdot 4n + 2n]$ is continuous, for an integer $i > 0$. A continuous cluster of length at least $2n$ necessarily contains all $m$ processors. Finally, Procedure Cluster-Synch is executed causing all $m$ processors to synchronize. This completes the description of the Procedure. The pseudo-code of the procedure is given below.

[1]Actually, at time $t_0$, it is sufficient for each processor to turn the radio on for a single time unit. The figure reflects the description of the algorithm in which at time $t_0$ the entire $k$-basic policy is performed. (For analysis purposes.)

**Algorithm 5** Procedure Synchronize()

An algorithm for a processor $v$

$k = \left\lceil \sqrt{8 \cdot n/m} \right\rceil$
$\tau = 0$
**for** $i = 1, 2, ..., \lceil \log n \rceil$ **do**
  $J$ := Cluster-Synch($\tau$, $k$)
  $\tau$ := Flatten($J, k$)
**end for**
Cluster-Synch($\tau$, $k$)

Procedure Synchronize preserves cluster distances in each phase in the following sense. Suppose that two processors $u$ and $v$ wake up at global times $t_u$ and $t_v$ respectively. Then, for $i = 1, 2, ..., \log n$, there are clusters $c_i$ and $c'_i$ such that $c_i$ covers an interval containing the point $(t_u + 4n \cdot i)$, and $c'_i$ covers an interval containing the point $(t_v + 4n \cdot i)$. Moreover, the cluster $c_i$ contains the processor $u$, and the cluster $c'_i$ contains the processor $v$. This observation, which is a consequence of Lemma 3.4, is summarized below.

**Corollary 3.5.** *Suppose that a processor $v$ performs the $k$-basic policy in time $t \in [0, 2n]$. If a cluster $c$ covers an interval containing the time point $(t + 4n \cdot i)$ for some integer $i > 0$, then $c$ contains $v$.*

In each phase of Procedure Synchronize, after the execution of Procedure Flatten, the sum of lengths of produced clusters is at least $k^2 \cdot m > 2n$. Consequently, at least two clusters overlap in each phase, and the number of clusters is decreased in each phase. Hence, it is obvious that $m$ phases are sufficient to merge all clusters into a single cluster. However, the merging process is actually much faster. The next Lemma states that after $\log n$ phases there is a single cluster containing all $m$ processors.

**Lemma 3.6.** *Once Procedure Synchronize is executed the global time interval $[\lceil \log n \rceil \cdot 4n, \lceil \log n \rceil \cdot 4n + 2n]$ is continuous.*

*Proof.* Suppose without loss of generality that the length of the $k$-basic policy $k + k^2$ satisfies $k + k^2 \leq n$. (Otherwise all processors overlap with the first awaking processor, and the problem becomes trivial.) In the execution of Procedure Synchronize all processors perform the $k$-basic policy completely during the interval $p_0 = [t_0, s_0] = [0, 2n]$. Hence, the covering-weight of the interval $p_0$ is at least $k^2 \cdot m \geq 8n$. The covering-density of the interval is at least 4. We define a series of intervals $p_1 = [t_1, s_1], p_2 = [t_2, s_2], ..., p_\lambda = [t_\lambda, s_\lambda]$ as follows. For $i = 0, 2, ..., \lambda - 1$, if $cden([t_i, \lceil \frac{1}{2}(t_i + s_i) \rceil]) > cden([\lceil \frac{1}{2}(t_i + s_i) \rceil, s_i])$ then $p_{i+1} = [t_i, \lceil \frac{1}{2}(t_i + s_i) \rceil]$. Otherwise $p_{i+1} = [\lceil \frac{1}{2}(t_i + s_i) \rceil, s_i]$. Observe that $p_{i+1}$ is contained in $p_i$. For $i = 1, 2, ..., \lambda$, the covering-density of $p_i$ is at least 4. See Figure 4 below for an illustration. Next, we define another interval series $p'_0, p'_1, ..., p'_\lambda$ as follows. $p'_0 = p_\lambda$, and for $i = 1, 2, .., \lambda$, $p'_i = [t'_i, s'_i] = [t_{\lambda-i} + i \cdot 4n, s_{\lambda-i} + i \cdot 4n]$.

Set $\lambda = \lceil \log n \rceil$. We prove by induction on $i$ that $p'_i$ is continuous, for $i = 1, 2, ..., \lambda$.

**Base (i = 1)**: Observe that the length of $p'_0$ is $len(p'_0) = len(p_\lambda) \leq 2$. (Since $len(p_i) \leq n/2^i + 1$.) Also $cden(p_\lambda) \geq 4$. Hence, by Lemma 3.4 (3), once the clusters of the first phase are flattened the interval $[4n + t_\lambda - len(p_\lambda), 4n + s_\lambda + len(p_\lambda)]$ is continuous. Hence, the interval $p'_1$ is continuous.

**induction step:** By induction hypothesis, assume that $p'_{i-1} = [t_{\lambda-i+1} + (i-1) \cdot 4n, s_{\lambda-i+1} + (i-1) \cdot 4n]$ is continuous. Thus, there is a cluster $c$ that covers an interval containing $p'_{i-1}$. By Corollary 3.5 the processors of clusters covering intervals that intersect with $p_{\lambda-i+1}$ are contained in $c$. Suppose that there are $\ell$ processors in $c$. Since the covering-density of $p_{\lambda-i+1}$ is at least 4, it holds that $\ell \cdot k^2 \geq 4 \cdot len(p_{\lambda-i+1}) = 4 \cdot len(p'_{i-1})$. Hence, $\ell \cdot k^2 \geq 4 \cdot (s_{\lambda-i+1} - t_{\lambda-i+1}) = 4 \cdot (s'_{i-1} - t'_{i-1})$. By Lemma 3.4 (3), once the clusters of the phase $i$ are flattened the interval $[i \cdot 4n + t_{\lambda-i+1} - len(p_{\lambda-i+1}), i \cdot 4n + s_{\lambda-i+1} + len(p_{\lambda-i+1})]$ is continuous. Hence, the interval $p'_i = [t_{\lambda-i} + i \cdot 4n, s_{\lambda-i} + i \cdot 4n]$ is continuous. □

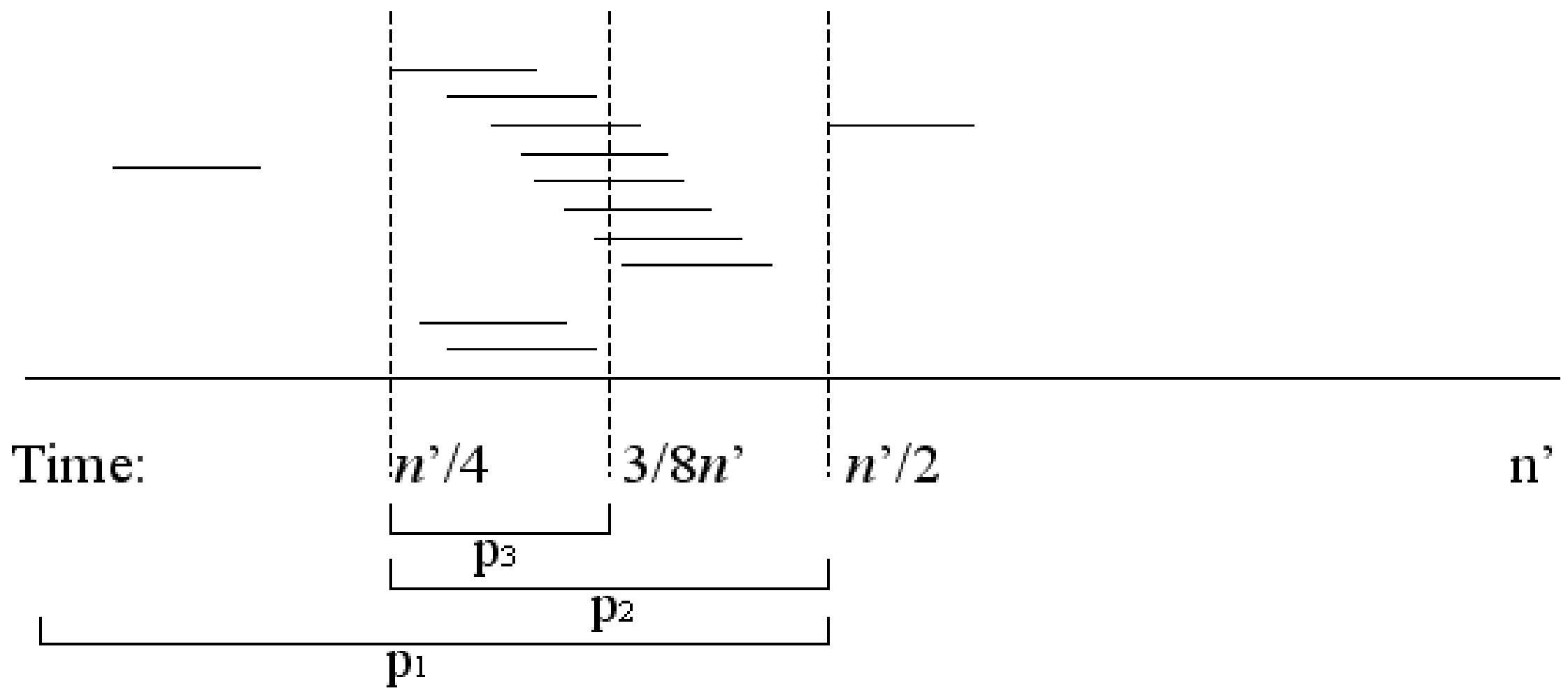

**Fig. 4.** *Illustration of the intervals $p_i$. It holds that $n' = 2n$.*

By Corollary 3.5 and Lemma 3.6, all $m$ processors are synchronized during global time interval $[\lceil \log n \rceil \cdot 4n, \lceil \log n \rceil \cdot 4n + 2n]$. Each processor performs the $k$-basic policy a constant number of times in each phase. Hence, in each phase, the number of time units in which each processor turns its radio on is $O(k) = O(\sqrt{n/m})$. The properties of Procedure Synchronize are summarized in the following Theorem.

**Theorem 3.7.** *Procedure Synchronize performs clock synchronization of $m$ processors waking up at arbitrary time points from the interval $[0, n]$. The energy efficiency of each processor is $O(\sqrt{n/m} \cdot \log n)$.*

### 3.2 Procedure Dynamic-Synch

In this section we show that by using a more sophisticated procedures one can achieve energy efficiency of $O(\sqrt{n/m})$ per processor. We start with describing a gas-stations riddle whose solution gives an intuition to the main ideas of the procedures we devise in this section. $m$ gas stations are arbitrarily placed on a one-way circular road. The total amount of gas in all stations is sufficient for a car to complete exactly two laps on the road. The car's gas-tank is sufficiently large, hence, each time the car approaches a station, it can add all the gas of the station to its tank. Can a car with an initially empty gas-tank start from one of the stations, and complete an entire lap? The answer to this riddle is affirmative. There always exists such a station. To find the station, select an arbitrarily station $p$ on the road. Place the car at the earliest station $s \neq p$ before $p$ (with respect to the driving direction) such that the car is able to arrive from $s$ to $p$. In other words, if the car is placed at a station that appears before $s$ it gets out of gas before arriving to $s$. (If there is no such station, then a car placed at station $p$ can complete an entire lap, and we are done.) The car drives form $s$ to $p$. When it arrives to $p$ it has enough gas to complete an entire lap, since the gas stations in the interval from $p$ to $s$ have not enough fuel to complete this interval. Consequently, the fuel of the stations in the interval from $s$ to $p$ is sufficient for completing this interval plus another complete lap.

In our algorithms the stations represent processors. Using the fuel of a certain station represents turning the radio on by the appropriate processor. However, using $x$ units of fuel represent a radio use of $O(\sqrt{x})$. For $i = 1, 2, ..., m$, the goal of a processor $i$ is to execute its radio use policy in the time interval in which the car would use the gas of station $i$. Since in each time unit the car uses the gas from only one station, there are no time units in which more than one main part of a policy is executed. However, for a processor to be able to determine the appropriate intervals a more sophisticated flattening procedure has to be used.

We devise a procedure called *Dynamic Flattening.* The use of dynamic flattening allows completing the synchronization in two phases instead of $O(\log n)$ phases that are required by Procedure Synchronize that was devised in the previous section. The main difference of Procedure Dynamic Flattening comparing to Procedure Flatten is that the scheduling stage is performed during the first execution of the policy rather than in the end of a phase. This scheduling is performed only once, shortly after a processor wakes up. A processor schedules the next execution of its policy to the first available free interval, i.e., an interval in which no other processor is scheduled. To this end, a queue of processors is maintained by each cluster. Consequently, the next policy execution of each processor $v$ is scheduled in such a way that the main part of $v$'s policy does not overlap with any of the other $m-1$ processors when they execute the main parts of their policies after scheduling. (In contrast, in Procedure Flatten the new scheduling of phase $i$ guarantees only that the main part of the policy of $v$ does not overlap with any processor in the cluster containing $v$ in phase $i$.) At time $2n$ at least $\frac{1}{2}m$ processors are scheduled one after the other to perform their policy. As a result, the global interval $[2n, 4n]$ is continuous. To guarantee that all $m$ processor perform their policy during this interval, each processor perform an additional independent invocation of its policy at time $2n$ from wake up.

The algorithm that employs this idea is called *Procedure Dynamic-Synch.*

**Informal description of Procedure Dynamic-Syncn (for each processor $v \in V$)**

**step 1.** The vertex $v$ sets $k := \left\lceil \sqrt{8 \cdot n/m} \right\rceil$, and performs the initial part of the $k$-basic policy.

**step 2.** If during step 1 one of the following holds: (i) $v$ does not discover any other processor whose radio is turned on, or (ii) all discovered processors have waken up after $v$ did, or have waken up at the same time as $v$ but have smaller Ids than that of $v$, then $v$ initializes a cluster $c$ and an empty queue $q_c$. The processor $v$ enqueues itself on $q_c$ and starts the main part of its policy once the initial part is complete.

**step 3 (Dynamic Flattening)** Otherwise, a queue $q$ is already initialized and maintained by the processor $u$ currently executing the main part of its policy. (The queue $q$ was created by the earliest processor in the cluster and passed in a rely-race manner. We stress that $u$ is not necessarily the earliest processor in the cluster.) Then $v$ enqueues itself on $q$ by communicating with $u$, and receives the number $\ell$ of processors that appear in $q$ before $v$. Suppose that $u$ has performed the main part of its policy for $r$ rounds once communicating with $v$. Then $v$ schedules the next $k$-basic policy execution such that the main part of its policy is executed in $(\ell - 1) \cdot k^2 - r$ time units. Such scheduling guarantees that policies of processors in $q$ are executed one after the other immediately, in the order they appear in $q$.

**step 4.** Once a processor completes executing its main part it dequeues itself from $q$ and passes $q$ to the next scheduled processor (with which it necessarily overlaps).

**step 5.** Execute the $k$-basic policy at time $2n + 1$ from wake up. (Independently of steps 1-4.)

This completes the description of the procedure. Its formal description and pseudocode are given in Appendix A. Its properties are summarized below. (See Figure 5 for an illustration.)

**Lemma 3.8.** *Suppose that $m$ processors wakeup during the global time interval $[0, n]$, and execute procedure Dynamic-Synch. Then the following hold: (1) for any pair of processors $u$ and $v$, their main parts are executed in distinct time intervals (that have no common time points) in the global interval $[0, 2n]$, (2) each cluster $c$ that covers an interval in $[0, 2n]$ satisfies that $cden(c) \leq 1$,*
*(3) there exists a cluster $c'$ that covers an interval containing the global time point $2n$. At global time $2n$*

*the queue of $c'$ contains at least $m/2$ processors.*

*Proof.* (1) Let $c$ be the dynamic cluster formed by Procedure Dynamic-Synch that contains $u$. Let $t_c$ be the global time of wake up of the earliest processor $w$ in $c$. (The time $t_c$ is also the start point of the original cluster of $w$.) Suppose that a processor $u$ has $\ell_u$ processors in $c$ that wake up before $u$, or wake up at the same time as $u$ but have greater $Id$s. Procedure Dynamic-Flattening schedules the main part of $u$ to be executed in the interval $I_u = (t_c + \ell_u \cdot k^2,\ t_c + (\ell_u + 1) \cdot k^2]$. If a processor $v \neq u$ belongs to $c$ then $\ell_v \neq \ell_u$. The interval in which the main part of $v$ is executed is $I_v = (t_c + \ell_v \cdot k^2,\ t_c + (\ell_v + 1) \cdot k^2]$. Hence, it holds that $I_u \cap I_v = \emptyset$. If $v$ does not belong to $c$, then by definition $I_u$ and $I_v$ do not have common time points.
(2) Consider a cluster $c$ that covers an interval $p = [s, t]$ in $[0, 2n]$. By (1) the main parts of policies in $c$ occur in distinct time intervals. Consequently, the sum of lengths of the main parts is at most $len(p)$. Hence $cden(c) \leq 1$.
(3) At global time point $2n$ all processors have already waken up and entered queues of leaders. By (2) at most $m/2$ processors performed the main part in the interval $[0, 2n]$. (Because $k^2 \cdot m/2 > 2n$. Thus, if more than $m/2$ processors perform the main parts in the interval $[0, 2n]$, at least one cluster must have density greater than 1. A contradiction.) Therefore there exist a cluster $c'$ covering an interval containing the point $2n$. At time point $2n$ the queue of the current leader in $c'$ contains all processors that have not executed the main part before time $2n$. Hence, it contains at least $m/2$ processors. □

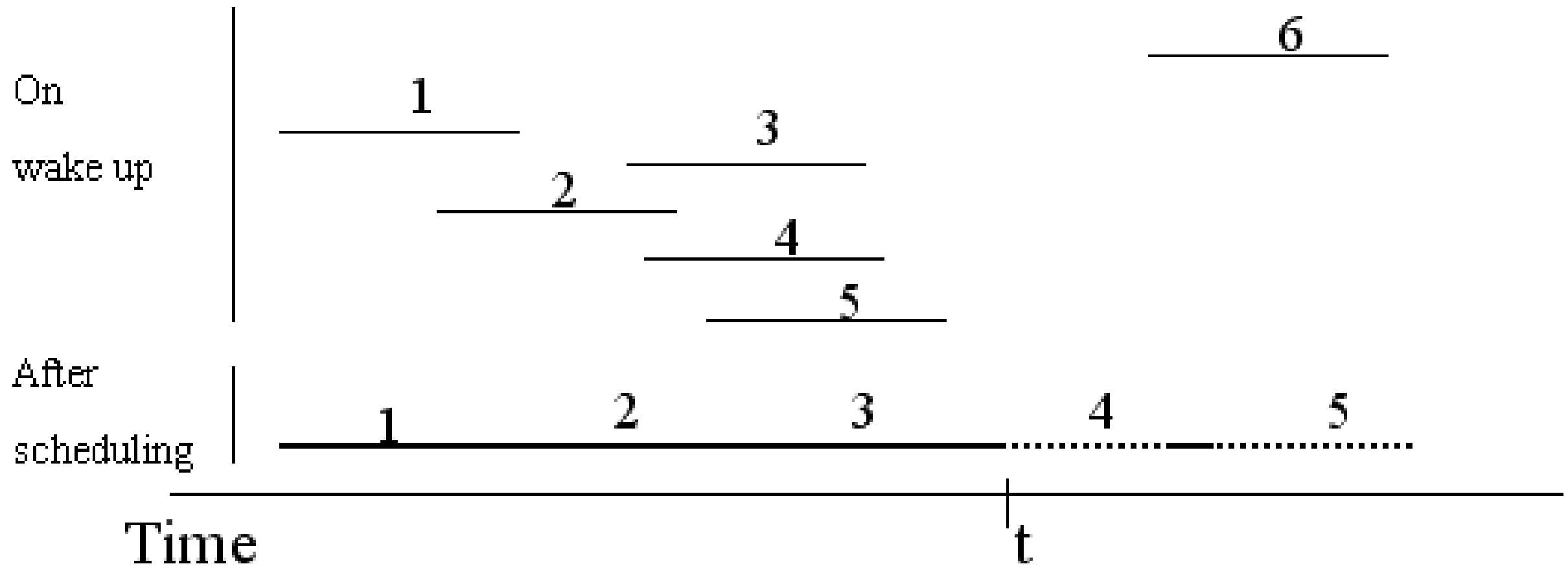

**Fig. 5.** *Illustration of Procedure Dynamic-Sync at time point $t$. Processors 1,2,3 have already performed the scheduled main part of their policy. Processors 4,5 are already scheduled, but have not performed their scheduled main part yet. Processor 6 will be scheduled once the main part of processor 5 is performed.*

By Lemma 3.8 (3), at global time point $2n$ at least $m/2$ processors are scheduled consequently. Hence the global interval $[2n, 4n]$ is continuous. All m processors execute their policy during this interval. Therefore all $m$ processors synchronize their clocks. Each processor execute the $k$-basic policy (fully or partly) 3 times. The correctness of procedure Dynamic-synch is summarized in the next Theorem.

**Theorem 3.9.** *Procedure Dynamic-synch synchronizes the clock of $m$ processors that wake up in the interval $[0, n]$. The energy efficiency per processor is $O(\sqrt{n/m})$.*

# 4 Lower Bounds

In this section we show strong lower bounds for energy use in clock synchronization in *general graphs*. We consider two scenarios. In the first scenario the energy efficiency of an algorithm is the maximum energy efficiency of a processor. This is the scenario discussed in previous sections. In the second scenario the energy efficiency of an algorithm is the average of energy consumed by the processors in the worst case. Observe that the second scenario is weaker in the sense that an algorithm with energy efficiency $O(k)$ in the second scenario may have energy efficiency $\omega(k)$ in the first scenario. The goal of an efficient algorithm in the first scenario is minimizing the maximum radio use of a processor. On the other hand, the goal in the second scenario is minimizing the sum of energy used by all processors. We prove our lower bounds for both scenarios. Moreover, our lower bounds apply not only to general graphs but also to specific families of graphs that are used to model wireless networks, such as unit disk graphs. We require the following results from [2].

**Lemma 4.1.** *[2] Suppose that each processor $v_1, v_2, ..., v_m$ in the complete graph of $m$ processors turns its radio on for $o(\sqrt{n/m})$ time units. Then for any deterministic synchronization algorithm $\mathcal{A}$ there are global time points $t_1, t_2, ..., t_m \in [0, n]$ of wake up and execution of $\mathcal{A}$ by $v_1, v_2, ..., v_m$, respectively, such that no two processors overlap.*

**Lemma 4.2.** *[2] In a two-processor network, for any radio use policy used by two processors $u$ and $v$, if $u$ and $v$ turn their radio on for $o(\sqrt{n})$ times each, there exist waking up global times $t_u$,$t_v \in [0, n]$ of $u$ and $v$ respectively, such that $u$ and $v$ do not overlap.*

We start with considering the first scenario in which the energy efficiency of the algorithm is the maximum energy efficiency of a processor. Lemma 4.2 implies that a synchronization of a two-processor network has energy efficiency $\Omega(\sqrt{n})$. Consequently, a synchronization of any $m$-vertex network that contains an isolated vertex $w$ (a vertex with degree 1) has radio efficiency $\Omega(\sqrt{n})$ per processor. Otherwise, if all processors have radio efficiency $o(\sqrt{n})$, then there are global time points $t_w$, $t'_w \in [0, n]$ such that $w$ wakes up at time $t_w$, the neighbor of $w$ wakes up at time $t'_w$, and $w$ does not synchronize with its neighbor. Hence, if the goal is minimizing the maximum radio use per processor then any algorithm for general graphs has efficiency $\Omega(\sqrt{n})$ per processor.

Next, we consider the second scenario in which the energy efficiency of the algorithm is the average of energy consumed by the processors. Surprisingly, we get the same result even for this weaker scenario. Let $G' = (V', E')$ and $G'' = (V'', E'')$ be complete graphs of $m' = m'' = m/2$ vertices each. Suppose for contradiction that there exist a synchronization algorithm $\mathcal{A}$ for general graph of $m$ processors in which the sum of radio use of all processors is $o(m\sqrt{n})$. Then in any invocation of $\mathcal{A}$ on a graph $G = (V, E)$, there is a processor $v \in V$ whose radio use is $o(\sqrt{n})$. Suppose that all processors of $G'$ wake up at global time $t'$, and all processors of $G''$ wake up at global time $t''$. Let $X'$ denote an execution of $\mathcal{A}$ on $G'$, and $X''$ the execution of $\mathcal{A}$ on $G''$. There is a vertex $v' \in V'$ (respectively $v'' \in V''$) whose radio use in the execution $X'$ (resp., $X''$) is $o(\sqrt{n})$.

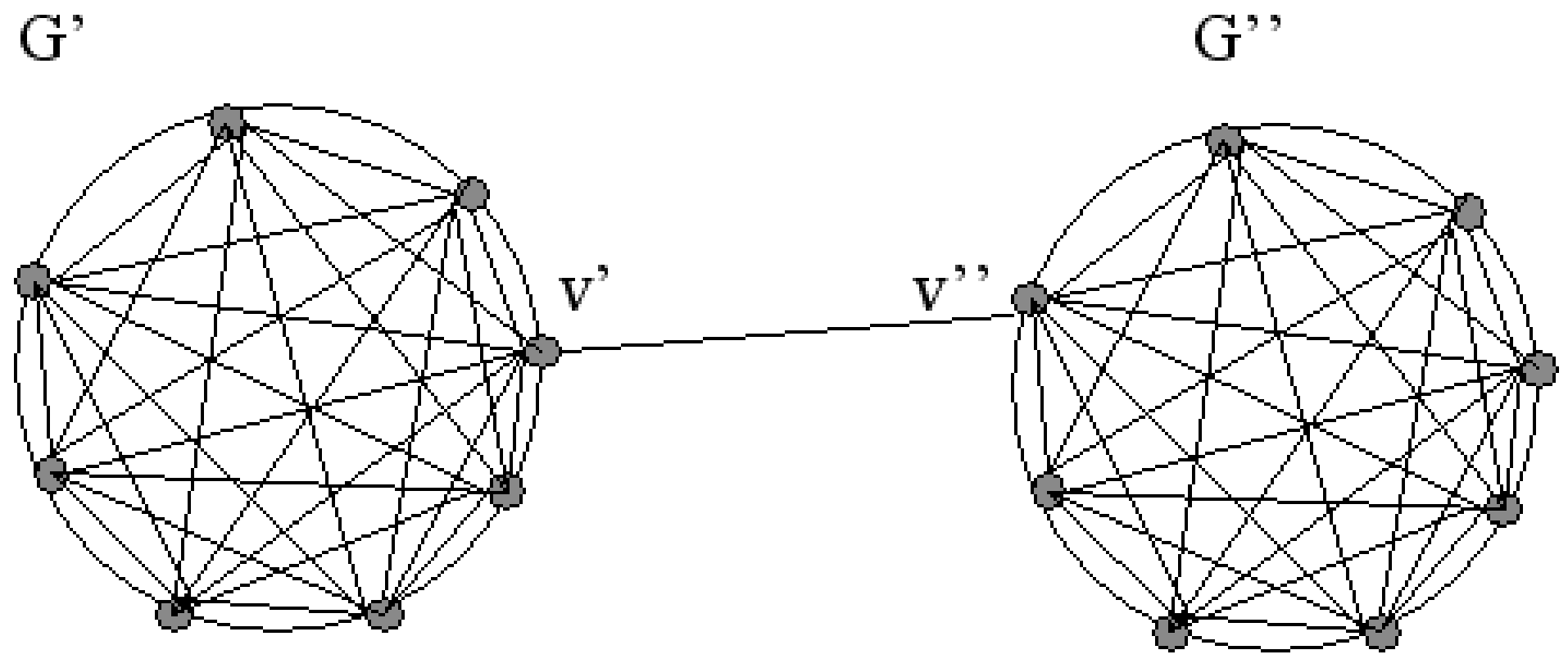

**Fig. 6.** *The graph $\hat{G}$ is obtained by connecting the vertex $v'$ of $G'$ and the vertex $v''$ of $G''$.*

Consider the graph $\hat{G} = (V' \cup V'', E' \cup E'' \cup (v', v''))$ that is achieved from $G'$ and $G''$ by connecting the vertices $v'$ and $v''$. (See Figure 6.) For $i, j \in \{0, 1, .., n\}$ Let $X(i, j)$ be the execution of $\mathcal{A}$ on $\hat{G}$ where all processors of $V'$ wake up at time $i$, and all processors of $V''$ wake up at time $j$. Since $\mathcal{A}$ synchronizes all processors of $\hat{G}$, the processors $v'$ and $v''$ overlap in each execution $X(i, j)$. Consider a two-vertex network $H$ consisting of a connected pair of vertices $u$,$w$. The vertex $u$ simulates the graph $G'$, and $w$ simulates $G''$. The vertex $u$ (respectively, $w$) turns its radio on if and only if $v'$ (resp., $v''$) turns its radio on. Once an algorithm $\mathcal{A}$ is invoked by $u$ (respectively, $w$) it simulates locally the execution of $\mathcal{A}$ on $G'$ (respectively, $G''$). For any execution $X(i, j)$ on $H$, the processors $u$ and $w$ overlap, since $v'$ and $v''$ overlap in the execution of $\mathcal{A}$ on $\hat{G}$. In each execution, each of the processors $u$ and $w$ has a radio use of $o(\sqrt{n})$. This is a contradiction to Lemma 4.2. Hence, at least $m/2$ vertices in $\hat{G}$ (either all vertices of $G'$ or all vertices of $G''$) must have radio efficiency $\Omega(\sqrt{n})$. We summarize this discussion in the following theorem.

**Theorem 4.3.** *In any clock synchronization algorithm $\mathcal{A}$ for general graphs the sum of processors radio use is $\Omega(m \cdot \sqrt{n})$. The energy efficiency of $\mathcal{A}$ in both scenarios is $\Omega(\sqrt{n})$.*

Observe that the construction described above applies also to *unit disk graphs*, i.e, graphs in which all vertices are placed in the plane, and have the same transmission range. Specifically, let $r$ be the radius of transmission in a unit disk graph. Place the vertices of $V'$ on the border of a cycle $c'$ of radius $1/2 \cdot r$. Similarly, place the vertices of $V''$ on the border of a cycle $c''$ of radius $1/2 \cdot r$. Place $v$ and $v'$ in distance $r$ one from the other, such that all other vertices $u' \in V'$, $u'' \in V''$ are in distance greater than $r$ one from the other. The Lower bound in Theorem 4.3 applies for this construction as well.

**Lemma 4.4.** *In any clock synchronization algorithm $\mathcal{A}$ for* unit disk graphs *the sum of processors radio use is $\Omega(m \cdot \sqrt{n})$. The energy efficiency of $\mathcal{A}$ in both scenarios is $\Omega(\sqrt{n})$.*

Next, we devise a lower bound for yet narrower family of graphs. An *$\ell$-connected graph* is a graph in which there are at least $\ell$ edge-disjoint paths connecting any pair of vertices. Consider an $m = 2m'$ vertex graph $\hat{G}$ consisting of two complete graphs $G' = (V' = \{v'_1, v'_2, ..., v'_{m/2}\}, E')$ and $G'' = (V'' = \{v''_1, v''_2, ..., v''_{m/2}\}, E'')$. Let $\ell$ be a positive integer parameter such that $\ell < m/4 - 2$. For $i = 1, 2, ..., \ell + 2$, the vertices $v'_i$ and $v''_i$ are connected. For $i, j > \ell + 2$, the vertices $v'_i$ and $v''_j$ are not connected. It is easy to see that $\hat{G}$ is an $\ell$-connected graph. (See Figure 7 below.) Suppose that each vertex $v'_1, v'_2, ..., v'_\ell, v''_1, v''_2, ...v''_\ell$ turns its radio on for $o(\sqrt{n/\ell})$ time units. Then, by Lemma 4.1, for any synchronization algorithm, there are time points such that no two processors among $v'_1, v'_2, ..., v'_\ell, v''_1, v''_2, ...v''_\ell$ overlap. Since the endpoints of each edge that connects $G'$ and $G''$ belong to $\{v'_1, v'_2, ..., v'_\ell, v''_1, v''_2, ...v''_\ell\}$, the network is not synchronized. Thus, if there are at least $\ell$ vertices in $G'$ that have radio use $o(\sqrt{n/\ell})$ each, and at least $\ell$ vertices in $G''$ that have radio use $o(\sqrt{n/\ell})$ each, the network is not synchronized. Consequently, at least $m - 2\ell + 1 = \Omega(m)$ vertices must use the radio for $\Omega(\sqrt{n/\ell})$ time units each in order to synchronize the network. This result is stated in the following theorem.

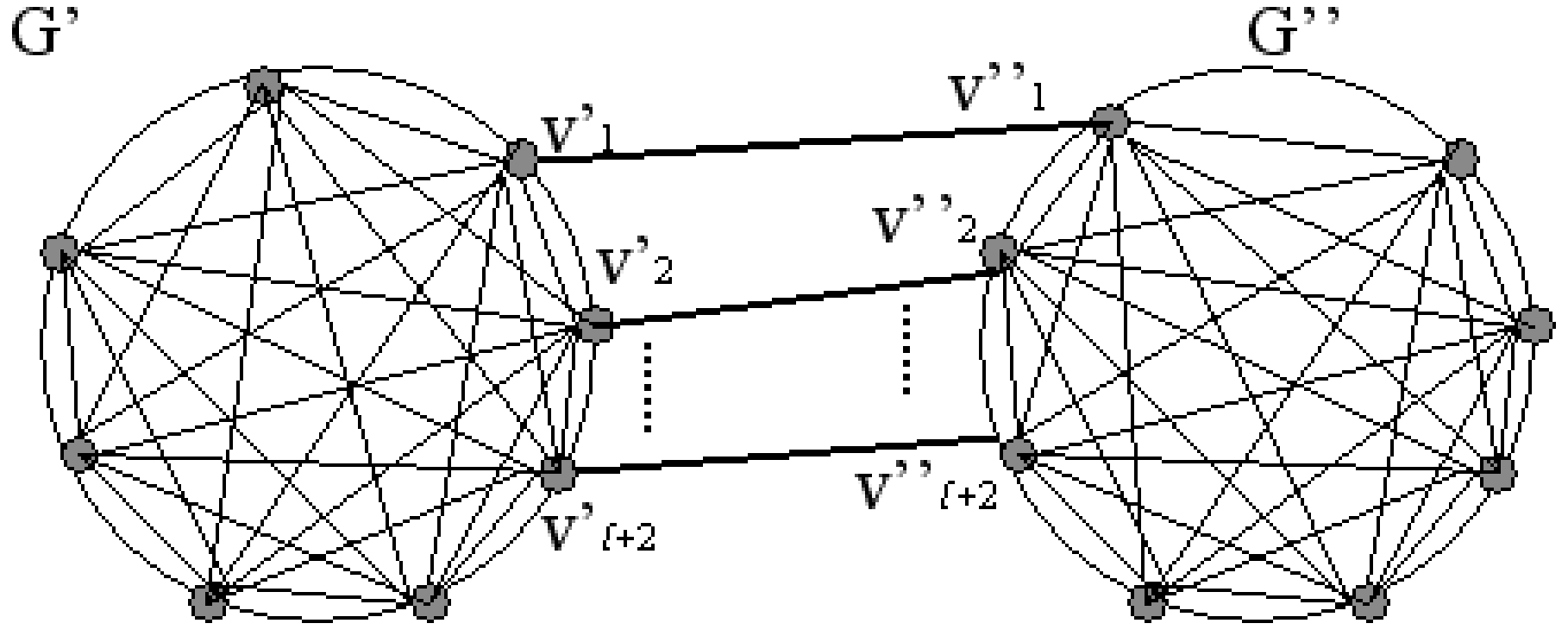

**Fig. 7.** *The graph $\hat{G}$ is obtained by connecting the vertex $v'_i$ of $G'$ and the vertex $v''_i$ of $G''$, for $i = 1, 2, ..., \ell + 2$.*

**Theorem 4.5.** *For a positive integer parameter $\ell < m/4 - 2$, in any clock synchronization algorithm $\mathcal{A}$ for $\ell$-connected graphs the sum of processors radio use is $\Omega(m \cdot \sqrt{n/\ell})$. The energy efficiency of $\mathcal{A}$ in both scenarios is $\Omega(\sqrt{n/\ell})$.*

## 5 Conclusion

In this paper we have devised optimal radio-use deterministic algorithms for clock synchronization in single-hop networks with energy efficiency $\Theta(\sqrt{n/m})$. We also proved lower bounds of $\Omega(\sqrt{n})$ for multi-hop networks. Our results suggest that in order to beat this bound of $\Omega(\sqrt{n})$, each neighborhood in the graph must be highly connected, containing no isolated regions. For wireless networks, this requires a certain level of uniformity in the processors distribution. In other words, for each processor $u$, each neighbor of $u$ must be in the communication range of a significant number of other neighbors of $u$.

In [2] a deterministic synchronization algorithm was devised for two-processor networks with efficiency $O(\sqrt{n})$. This algorithm can be used also in multi-hop network in order to synchronize each processor with its neighbors. The energy efficiency in this case is $O(\sqrt{n})$ per processors. Somewhat surprisingly, our lower bounds imply that this simple approach is optimal in general multi-hop networks.

## References

[1] P. Blum, L. Meier and L. Thiele Improved interval-based clock synchronization in sensor networks IPSN '04: Proceedings of the third international symposium on Information processing in sensor networks, pp. 349-358, 2004.

[2] M. Bradonjic, E. Kohler, and R. Ostrovsky Near-Optimal Radio Use For Wireless Network Synchronization. In proc. of the 5th International Workshop on Algorithmic Aspects of Wireless Sensor Networks, ALGOSENSORS, pp. 15-28, 2009.

[3] A. Boulis, M. Srivastava Node-Level Energy Management for Sensor Networks in the Presence of Multiple Applications. Wireless Networks 10(6): 737-746 (2004)

[4] A. Boulis, S. Ganeriwal, M. Srivastava Aggregation in sensor networks: an energy-accuracy trade-off. Ad Hoc Networks 1(2-3): 317-331 (2003)

[5] S.F. Bush Low-energy sensor network time synchronization as an emergent property. In Proc. of the 14th International Conference on Communications and Networks, ICCCN, pp. 93-98, 2005.

[6] B. Chlebus and L. Gasieniec and A. Gibbons and A. Pelc and W. Rytter Deterministic broadcasting in ad hoc radio networks Distributed Computing, Volume 15, Number 1, January 2002 Pages: 27 - 38

[7] J. Elson and K. Römer Wireless sensor networks: a new regime for time synchronization SIGCOMM Comput. Commun. Rev., vol. 33. no. 1. pp. 149-154, 2003.

[8] J. Elson, L. Girod, and D. Estrin Fine-Grained Network Time Synchronization using Reference Broadcasts. Proc. Fifth Symposium on Operating Systems Design and Implementation (OSDI 2002), Vol 36, pp. 147163, 2002.

[9] J. Elson and K. Römer Wireless sensor networks: a new regime for time synchronization. Proceedings of the First Workshop on Hot Topics In Networks (HotNets-I), Princeton, New Jersey. October 28-29 2002.

[10] R. Fan, I. Chakraborty, N. Lynch Clock Synchronization for Wireless Networks. OPODIS 2004: pp. 400-414.

[11] N. Honda and Y. Nishitani The Firing Squad Synchronization Problem for Graphs. Theoretical Computer Sciences 14(1):39-61, April 1981.

[12] A. Kesselman and D. Kowalski Fast distributed algorithm for convergecast in ad hoc geometric radio networks. Conference onWireless On demand Network Systems and Services, 2005.

[13] K. Kobayashi The Firing squad synchronization problem for a class of polyautomata networks. Journal of Computer and System Science 17:300-318, 1978.

[14] K.Kothapalli, M. Onus, A. Richa and C. Scheideler Efficient Broadcasting and Gathering in Wireless Ad Hoc Networks in IEEE International Symposium on Parallel Architectures, Algorithms and Networks(ISPAN), 2005.

[15] D. Kowalski and A. Pelc Broadcasting in undirected ad hoc radio networks. In Proceedings of the twenty-second annual symposium on Principles of distributed computing, pages 73-82. ACM Press, 2003.

[16] D. Kowalski and A. Pelc Faster deterministic broadcasting in ad hoc radio networks. Proc. 20th Ann. Symp. on Theor. Aspects of Comp. Sci. (STACS'2003), LNCS 2607, 109-120, 2003.

[17] H. Kopetz, W. Ochsenreiter Global time in distributed real-time systems. Technical Report 15/89, Technische Universitat Wien, Wien Austria (1989).

[18] C. Lenzen, T. Locher, P. Sommer, R. Wattenhofer Clock Synchronization: Open Problems in Theory and Practice. SOFSEM 2010. 36th International Conference on Current Trends in Theory and Practice of Computer Science. Proceedings, pp. 61-70 , January 2010.

[19] D.L. Mills Internet time synchronization: the network time protocol. Communications, IEEE Transactions on , vol.39, no.10, pp.1482-1493, Oct 1991.

[20] T. Moscibroda, P. Von Rickenbach, R. Wattenhofer Analyzing the Energy-Latency Trade-Off During the Deployment of Sensor Networks. INFOCOM 2006. 25th IEEE International Conference on Computer Communications. Proceedings, pp.1-13, April 2006.

[21] M. McGlynn and S. Borbash Birthday protocols for low energy deployment and flexible neighbor discovery in ad hoc wireless networks. MobiHoc '01: Proceedings of the 2nd ACM international symposium on Mobile ad hoc networking & computing, pp. 137–145, 2001.

[22] V. Park, M. Corson A Highly Adaptive Distributed Routing Algorithm for Mobile Wireless Networks INFOCOM '97. Sixteenth Annual Joint Conference of the IEEE Computer and Communications Societies. Driving the Information Revolution, 1997.

[23] S. Palchaudhuri and D. Johnson Birthday paradox for energy conservation in sensor networks. Proceedings of the 5th Symposium of Operating Systems Design and Implementation, 2002.

[24] J. Polastre, J. Hill, and D. Culler Versatile low power media access for wireless sensor networks. In Proceedings of the 2nd international Conference on Embedded Networked Sensor Systems (Baltimore, MD, USA, November 03 - 05, 2004). SenSys '04. ACM Press, New York, NY, 95-107.

[25] M.L. Sichitiu, C. Veerarittiphan Simple, accurate time synchronization for wireless sensor networks Wireless Communications and Networking, 2003. WCNC 2003. 2003 IEEE, vol. 2, pp. 1266- 1273, 16-20 March 2003.

[26] V. Shnayder, M. Hempstead, B. Chen, G. Allen and M. Welsh Simulating the power consumption of large-scale sensor network applications. In Proceedings of the 2nd international conference on embedded networked sensor systems, SenSys '04, pp. 188–200, 2004.

[27] C. Schurgers, V. Raghunathan, M. Srivastava Power management for energy-aware communication systems. ACM Trans. Embedded Comput. Syst. 2(3): 431-447 (2003).

[28] F. Sivrikaya, B. Yener Time synchronization in sensor networks: a survey. Network, IEEE , vol.18, no.4, pp. 45-50, July-Aug. 2004.

[29] M. L. Sichitiu and C. Veerarittiphan Simple, Accurate Time Synchronization for Wireless Sensor Networks. Proc. IEEE Wireless Communications and Networking Conference (WCNC 2003), pp. 12661273, 2003.

[30] B. Sundararaman, U. Buy and A. D. Kshemkalyani Clock synchronization for wireless sensor networks: a survey. Ad-hoc Networks, 3(3): 281-323, May 2005.

# Appendix

## A Procedure Dynamic-Synch

The pseudocode of Procedure Dynamic-Synch is given below. Each vertex maintains the local variables *candidate*, *winner*, and $q$. During the execution of the main part by a processor $v$, the processor $v$ is called a *temporary leader*. The goal of the procedure is to guarantee that there is at most one leader at any time point. However, during the execution of the algorithm there may be numerous leaders, since different processors may become temporary leaders at distinct time intervals. To this end, each vertex $u$ initially sets its local variable *candidate* to true, i.e., it is a candidate for leadership. Then it sends an initial message for $k$ rounds. If $u$ receives a response from a leader, $u$ cannot become a leader in this phase. Hence, $u$ sets the variable *candidate* as false. If, on the other hand, $u$ does not receive a response from a leader during the initial phase (the first $k$ rounds), it can become a leader. However, additional processor may try doing so concurrently. In order to select exactly one leader at a time, a local variable *winner* is maintained by each processor. Similarly to leadership, winning is a temporary state. In other words, at any time point there is at most one winner, but in different time points there may be distinct winners. A processor $u$ is set as temporary winner only if in the first round performed by $u$ there are no other winners, or if all other potential winners have smaller Ids. (Consequently, these potential winners lose.)

A temporary winner candidate becomes a temporary leader once its initial part is complete. It sends a response for each initial message it receives from other processors. The response contains the information required for the other processor to schedule a time interval in which it can become a temporary leader and execute its main part exclusively. To this end, a temporary leader $u$ maintains a queue $q$, that initially contains only $Id(u)$. The queue $q$ represents all processors that are already scheduled to perform their main parts, but have not completed the main parts yet. Each received message with an Id of another processor $v$ is enqueued on $q$. A response is sent with the position of $Id(v)$ in $q$. Consequently, any two distinct processors receive from $u$ a distinct position, and schedule the executions of their main parts to distinct intervals. Moreover, once $u$ completes its main part it pops its Id from $q$, and passes $q$ to the next temporary leader that is scheduled right after $u$. Consequently, any two processors schedule distinct time intervals for their main parts, even if they communicate with different leaders.

The exact computation of the time interval to perform the main part is performed by procedure Dynamic-Flattening as described above. Its pseudocode follows. The procedure accepts as input the variables $k$, *candidate*, *winner*, $q$, $\ell$, $dif$. The variable $\ell$ is the position of the processor in the queue of the temporary leader. The variable $dif$ is the difference between the number of rounds the main part of the leader has executed, and the number of rounds that current processor has executed. Based on this information the processor schedules the time of execution of its main part, and becomes a temporary leader during this period.

**Algorithm 6** Procedure Dynamic-Synch()

An algorithm for a processor $v$. The rounds are counted from wakeup.

$k = \left\lceil \sqrt{8 \cdot n/m} \right\rceil$ ; $candidate := true$ ; $winner := true$ ; $q := \{Id(v)\}$
/*** initial part ***/
**for** rounds $r := 1, 2, ..., k$ **do**
  transmit the message initial$(Id(v), r)$
  **for** each received message initial$(Id(u), r_u)$ **do**
    /* local processing of messages is by ascending order of Ids */
    **if** $(r = 1)$ and $(r_u > r$ or $(r_u = r$ and $Id(u) > Id(v)))$ **then**
      $winner$ := false
    **end if**
    **if** $Id(u)$ is not in $q$ **then**
      $q$.enqueue$(Id(u))$ /* $q[|q| + 1] := Id(u)$ */
    **end if**
  **end for**
  **if** $candidate$ and received the message initial-response$(Id(v), \ell, \hat{r})$ **then**
    $candidate := false$
    Dynamic-Flattening($k$, $candidate$, $winner$, $q$, $\ell$ , $\hat{r} - r$)
  **end if**
  **if** $r = k$ and $candidate$ and $winner$ **then**
    **for** $j := 1, 2, ..., |q|$ **do**
      transmit the message initial-response$(q[j], j, 0)$
    **end for**
    Dynamic-Flattening($k$, $candidate$, $winner$, $q$, 0 , 0)
  **end if**
**end for**
execute the $k$-basic policy independently starting from round $2n + 1$

**Algorithm 7** Procedure Dynamic-Flattening($k$, $candidate$, $winner$, $q$, $\ell$, $dif$)

An algorithm for a processor $v$. The rounds are counted from wakeup.

**if** $candidate$ and $winner$ **then**
  $next := k$
**else**
  $next := (\ell - 1) \cdot k^2 - dif$
**end if**
/*** main part ***/
**for** rounds $r := next + k$, $next + 2k, ...,$ $next + k^2$ **do**
  **if** ($r = next + k$) and not ($candidate$ and $winner$) **then**
    receive the message pass($q'$)
    $q := q'$
  **end if**
  **for** each received message initial($Id(u), r_u$) **do**
    /* local processing of messages is by ascending order of Ids */
    **if** $Id(u)$ is not in $q$ **then**
      $q$.enqueue($Id(u)$)
    **end if**
  **end for**
  **for** $j := 1, 2, ..., |q|$ **do**
    transmit the message initial-response($q[j], j, r - next$)
  **end for**
**end for**
on round $next + k^2 + k$ perform q.dequeue() and transmit the message pass($q$)

## B General Scenarios

We begin with analyzing the scenario in which the shifts of wake up are not necessarily integer. In this scenario each processor $v$ wakes up at global time $t_v \in [0, n) \subseteq \mathcal{R}$, and all clocks proceed with the same speed. In this scenario it is impossible to achieve precise synchronization, since it is impossible to adjust a clock by an arbitrary small fraction of a unit. Therefore, the goal in this case is to set all clocks $\tau_1, \tau_2, ..., \tau_m$ such that at any time point the difference between $\tau_i$ and $\tau_j$ is at most 1, for $1 \leq i \neq j \leq m$.

Next, we describe an additional step that each processor must perform each time it updates its clock. Otherwise, there might be a difference between clocks that may grow significantly as multiple updates are performed. The additional step guarantees that the difference is always smaller than one time unit. Let $t$ be the maximum time required for a sent message to arrive at its destination. A time unit is set to $2t$. Consequently, if two processors turn their radio on, and overlap for at least half a unit, then they are able to communicate. A processor $u$ maintains an additional variable $q_u$ that holds a value from the range $[-\frac{1}{2}, \frac{1}{2}]$ that is initially set to 0. Each time a processor $v$ communicates with a processor $u$ it determines the length $q$ of the overlap, which is a fraction in the range $[\frac{1}{2}, 1]$. Next, it sets $q' := 1 - q$ if $u$ turned its radio on after $v$ did. Otherwise it sets $q' := q - 1$. Observe that $q'$ is a fraction in the range $[-\frac{1}{2}, \frac{1}{2}]$. It represents the time length between a clock tick of $u$, and a clock tick of $v$. Specifically, if $v$ increments its clock at global time $t$, then $u$ increments its clock at global time $t + q'$.

Suppose that a processor $u$ communicates with a processor $v$, and updates its clock $\tau_u$ to the clock $\tau_v$ of $v$. Then it should also update the variable $q_u$ as follows. $q_u := q_v + q'$. Consequently, the value of $q_u$ may rise beyond $\frac{1}{2}$, or fall below $-\frac{1}{2}$. In such cases, if $q_u > \frac{1}{2}$ set $\tau_u := \tau_u + 1$, and $q_u := q_u - 1$. If $q_u < -\frac{1}{2}$ set $\tau_u := \tau_u - 1$, and $q_u := q_u + 1$. This completes the description of the additional step. We show that processors that perform synchronization using this step are indeed synchronized.

**Lemma B.1.** *Suppose that processors $v_1, v_2, ..., v_\ell$ perform synchronization with the step described above. Then, at any time point after synchronization the difference between any two clock is at most 1.*

*Proof.* Suppose without loss of generality that $v_1$ is the earliest processor. We prove by induction on the number of processors that the difference between the clocks of $v_1$ and $v_j$ is at most $\frac{1}{2}$, for $j = 1, 2, ..., \ell$. The base case is trivial. For the induction step, Suppose that $j - 1$ processors $v_1, v_2, ..., v_{j-1}$ are synchronized with the earliest processor $v_1$. Observe that $q_{v_1} = 0$ since $v_1$ does not update its clock. Also, for $i = 1, 2, ..., j - 1$ it holds that $|\tau_{v_i} - \tau_{v_1}| \leq 1$, and $\tau_{v_i} + q_{v_i} = \tau_{v_1}$ where $-\frac{1}{2} \leq q_{v_i} \leq \frac{1}{2}$. Suppose also that an additional processor $u = v_j$ performs synchronization with a processor $v_k$, for any $1 \leq k \leq j - 1$. The processor $u$ sets $\tau_u := \tau_{v_k}$, and $q_u := q_{v_k} + q'$. At this point it is possible that the clocks of $v_1$ and $u$ have a difference larger than $\frac{1}{2}$, but is not greater than 1. However, once the step is completed, $q_u$ is updated in such a way that the difference becomes at most $\frac{1}{2}$, and it holds that $\tau_u + q_u = \tau_{v_1}$. □

Finally, we remark that the model is sufficiently expressive to capture an even more general case in which the clock speeds differ, as long as the ratio of different speeds is bounded by a constant. For full analysis see [2], Section 7.